\documentclass[twocolumn, epjc3]{svjour3}

\RequirePackage{graphicx} 
\usepackage{bm}
\usepackage{epsfig}
\usepackage{subfig} 
\usepackage{multirow}
\usepackage[colorlinks=true,citecolor=blue,filecolor=blue,linkcolor=blue,urlcolor=blue,pdftex]{hyperref}
\usepackage{mathtools}
\usepackage{wasysym}
\usepackage{supertabular,booktabs}
\usepackage[switch]{lineno} 
\usepackage[usenames,dvipsnames]{xcolor}
\usepackage[normalem]{ulem}
\usepackage{siunitx} 
\usepackage[symbol]{footmisc}

\journalname{Eur. Phys. J. C}

\begin{document}

\sloppy

\title{Sensitivity of the DARWIN observatory to the neutrinoless double beta decay of $^{136}$Xe}

\author{
F.~Agostini\thanksref{bologna} \and 
S.~E.~M.~Ahmed~Maouloud\thanksref{paris} \and 
L.~Althueser\thanksref{munster} \and
F.~Amaro\thanksref{coimbra} \and
B.~Antunovic\thanksref{vinca}\textsuperscript{,}\footnotemark[2] \and
E.~Aprile\thanksref{columbia} \and 
L.~Baudis\thanksref{zurich} \and
D.~Baur\thanksref{freiburg} \and
Y.~Biondi\thanksref{zurich} \and 
A.~Bismark\thanksref{zurich,freiburg} \and 
P.~A.~Breur\thanksref{nikhef}\textsuperscript{,}\footnotemark[3] \and
A.~Brown\thanksref{zurich} \and
G.~Bruno\thanksref{nyuad} \and
R.~Budnik\thanksref{wis} \and 
C.~Capelli\thanksref{zurich} \and
J.~Cardoso\thanksref{coimbra} \and 
D.~Cichon\thanksref{mpik} \and
M.~Clark\thanksref{purdue} \and
A.~P.~Colijn\thanksref{nikhef}\textsuperscript{,}\footnotemark[4] \and
J.~J.~Cuenca-Garc\'ia\thanksref{kit} \and
J.~P.~Cussonneau\thanksref{subatech} \and
M.~P.~Decowski\thanksref{nikhef} \and
A.~Depoian\thanksref{purdue} \and
J.~Dierle\thanksref{freiburg} \and
P.~Di~Gangi\thanksref{bologna} \and 
A.~Di~Giovanni\thanksref{nyuad} \and 
S.~Diglio\thanksref{subatech} \and
J.~M.~F.~dos~Santos\thanksref{coimbra} \and 
G.~Drexlin\thanksref{kit2} \and 
K.~Eitel\thanksref{kit} \and
R.~Engel\thanksref{kit} \and 
A.~D.~Ferella\thanksref{Univ_lAquila}\textsuperscript{,}\thanksref{LNGS_lAquila} \and
H.~Fischer\thanksref{freiburg} \and
M.~Galloway\thanksref{zurich} \and
F.~Gao\thanksref{columbia} \and 
F.~Girard\thanksref{zurich} \and
F.~Gl\"uck\thanksref{kit} \and 
L.~Grandi\thanksref{chicago} \and 
R.~Gr\"o\ss le\thanksref{kit} \and
R.~Gumbsheimer\thanksref{kit} \and 
S.~Hansmann-Menzemer\thanksref{heidelberguni} \and
F.~J\"org\thanksref{mpik} \and
G.~Khundzakishvili\thanksref{purdue} \and
A.~Kopec\thanksref{purdue} \and
F.~Kuger\thanksref{e1,freiburg} \and 
L.~M.~Krauss\thanksref{originsPF} \and
H.~Landsman\thanksref{wis} \and 
R.~F.~Lang\thanksref{purdue} \and 
S.~Lindemann\thanksref{freiburg} \and 
M.~Lindner\thanksref{mpik} \and 
J.~A.~M.~Lopes\thanksref{coimbra}\textsuperscript{,}\footnotemark[5] \and 
A.~Loya~Villalpando\thanksref{nikhef} \and
C.~Macolino\thanksref{lal} \and 
A.~Manfredini\thanksref{zurich} \and
T.~Marrod\'an~Undagoitia\thanksref{mpik} \and 
J.~Masbou\thanksref{subatech} \and
E.~Masson\thanksref{lal} \and 
P.~Meinhardt\thanksref{freiburg} \and
S.~Milutinovic\thanksref{vinca}\and 
A.~Molinario\thanksref{lngs} \and 
C.~M.~B.~Monteiro\thanksref{coimbra} \and 
M.~Murra\thanksref{munster} \and
U.~G.~Oberlack\thanksref{mainz} \and
M.~Pandurovic\thanksref{vinca}\and
R.~Peres\thanksref{zurich} \and 
J.~Pienaar\thanksref{chicago} \and 
M.~Pierre\thanksref{subatech} \and
V.~Pizzella\thanksref{mpik} \and
J.~Qin\thanksref{purdue} \and
D.~Ram\'irez~Garc\'ia\thanksref{freiburg} \and
S.~Reichard\thanksref{zurich} \and 
N.~Rupp\thanksref{mpik} \and 
P.~Sanchez-Lucas\thanksref{e2,zurich} \and 
G.~Sartorelli\thanksref{bologna} \and
D.~Schulte\thanksref{munster} \and
M.~Schumann\thanksref{freiburg} \and
L.~Scotto~Lavina\thanksref{paris} \and 
M.~Selvi\thanksref{bologna} \and
M.~Silva\thanksref{coimbra} \and 
H.~Simgen\thanksref{mpik} \and 
M.~Steidl\thanksref{kit} \and 
A.~Terliuk\thanksref{heidelberguni} \and
C.~Therreau\thanksref{subatech} \and
D.~Thers\thanksref{subatech} \and
K.~Thieme\thanksref{zurich} \and 
R.~Trotta\thanksref{imperial}\textsuperscript{,}\footnotemark[6] \and 
C.~D.~Tunnell\thanksref{rice} \and 
K.~Valerius\thanksref{kit} \and 
G.~Volta\thanksref{zurich} \and 
D.~Vorkapic\thanksref{vinca} \and 
C.~Weinheimer\thanksref{munster} \and
C.~Wittweg\thanksref{munster} \and
J.~Wolf\thanksref{kit2} \and 
J.~P.~Zopounidis\thanksref{paris} \and 
K.~Zuber\thanksref{dresden} 
(DARWIN~Collaboration\thanksref{e3})
}

\thankstext{e1}{Fabian.Kuger@physik.uni-freiburg.de}
\thankstext{e2}{patricia.sanchez@physik.uzh.ch}
\thankstext{e3}{darwin@lngs.infn.it}

\institute{Department of Physics and Astronomy, University of Bologna and INFN-Bologna, 40126 Bologna, Italy \label{bologna} \and 
LPNHE, Universit\'{e} Pierre et Marie Curie, Universit\'{e} Paris Diderot, CNRS/IN2P3, Paris 75252, France \label{paris} \and
Institut f\"ur Kernphysik, Westf\"alische Wilhelms-Universit\"at M\"unster, 48149 M\"unster, Germany \label{munster} \and
LIBPhys, Department of Physics, University of Coimbra, 3004-516 Coimbra, Portugal \label{coimbra} \and 
Vinca Institute of Nuclear Science, University of Belgrade, Mihajla Petrovica Alasa 12-14. Belgrade, Serbia \label{vinca} \and
Physics Department, Columbia University, New York, NY 10027, USA \label{columbia} \and  
Physik-Institut, University of Zurich, 8057  Zurich, Switzerland \label{zurich} \and
Physikalisches Institut, Universit\"at Freiburg, 79104 Freiburg, Germany \label{freiburg} \and
Nikhef and the University of Amsterdam, Science Park, 1098XG Amsterdam, Netherlands \label{nikhef}  \and
New York University Abu Dhabi, Abu Dhabi, United Arab Emirates \label{nyuad} \and
Department of Particle Physics and Astrophysics, Weizmann Institute of Science, Rehovot 7610001, Israel \label{wis} \and
Max-Planck-Institut f\"ur Kernphysik, 69117 Heidelberg, Germany \label{mpik}  \and
Department of Physics and Astronomy, Purdue University, West Lafayette, IN 47907, USA \label{purdue} \and
Institute for Nuclear Physics (IKP), Karlsruhe Institute of Technology (KIT), 76344 Eggenstein-Leopoldshafen, Germany \label{kit} \and
SUBATECH, IMT Atlantique, CNRS/IN2P3, Universit\'e de Nantes, Nantes 44307, France \label{subatech} \and
Institute of Experimental Particle Physics (ETP), Karlsruhe Institute of Technology (KIT), 76344 Eggenstein-Leopoldshafen, Germany \label{kit2} \and
Department of Physics and Chemistry, University of L’Aquila, 67100 L’Aquila, Italy \label{Univ_lAquila} \and
INFN-Laboratori Nazionali del Gran Sasso and Gran Sasso Science Institute, 67100 L’Aquila, Italy \label{LNGS_lAquila} \and
Department of Physics \& Kavli Institute for Cosmological Physics, University of Chicago, Chicago, IL 60637, USA \label{chicago} \and 
Physikalisches Institut, Ruprecht-Karls-Universit\"at Heidelberg, Heidelberg, Germany \label{heidelberguni} \and
The Origins Project Foundation, Phoenix, AZ 85020, USA \label{originsPF} \and
Universit\'e Paris-Saclay, CNRS/IN2P3, IJCLab, F-91405 Orsay, France \label{lal} \and
INFN-Laboratori Nazionali del Gran Sasso and Gran Sasso Science Institute, 67100 L'Aquila, Italy \label{lngs} \and
Institut f\"ur Physik \& Exzellenzcluster  PRISMA$^{+}$, Johannes Gutenberg-Universit\"at Mainz, 55099 Mainz, Germany \label{mainz} \and
Department of Physics, Imperial Centre for Inference and Cosmology, Imperial College London, London SW7 2AZ, UK \label{imperial} \and
Department of Physics and Astronomy, Rice University, Houston, TX 77005, USA \label{rice} \and
Institute for Nuclear and Particle Physics, TU Dresden, 01069 Dresden, Germany \label{dresden} 
}

\date{Received: date / Accepted: date}

\maketitle

\footnotetext[2]{Also at University of Banja Luka, Bosnia and Herzegovina}
\footnotetext[3]{Now at SLAC, Menlo Park, CA 94025, USA}
\footnotetext[4]{Also at Institute for Subatomic Physics, Utrecht University, Utrecht, Netherlands}
\footnotetext[5]{Also at Coimbra Polytechnic - ISEC, Coimbra, Portugal} 
\footnotetext[6]{Also at SISSA, Data Science Excellence Department, Trieste, Italy} 

\begin{abstract}

The DARWIN observatory is a proposed next-generation experiment to search for particle dark matter and for the neutrinoless double beta decay of $^{136}$Xe. 
{\color{black}Out of its \SI{50}{t} total natural xenon inventory, \SI{40}{t} will be the active target of a time projection chamber which thus contains} about \SI{3.6}{t} of $^{136}$Xe. Here, we show that its projected half-life sensitivity is \SI{2.4e27}{yr}, using a fiducial volume of \SI{5}{t} of natural xenon and \SI{10}{yr} of operation with a background rate of less than 0.2~events/(t~$\cdot$~yr) in the energy region of interest. This sensitivity is based on a detailed Monte Carlo simulation study of the background and event topologies in the large, homogeneous target. DARWIN will be comparable in its science reach to dedicated double beta decay experiments using xenon enriched in $^{136}$Xe.

\end{abstract}

\section{Introduction}
\label{sec:intro}

Neutrinos are the only known elementary particles that are Majorana fermion candidates, implying that they would be their own antiparticles. The most sensitive probe for the Majorana nature of neutrinos is an extremely rare nuclear decay process called neutrinoless double beta decay ($0\nu\beta\beta$), where a nucleus with mass number A and charge Z decays by emitting only two electrons and changes its charge by two units (A,Z)$\longrightarrow$(A,Z+2) + 2e$^{-}$. The observation of this decay would mean that lepton number is violated by two units and, {\color{black} in the standard light Majorana neutrino exchange scenario,} would yield information about the neutrino mass scale via the effective neutrino Majorana mass $\langle m_{\beta\beta}\rangle = |\Sigma_i U^{2}_{ei}m_i|$. The sum is over the neutrino mass eigenstates, $m_i$, and $U_{ei}$, the corresponding entries in  the lepton mixing matrix, which are complex numbers. The two-neutrino double beta decay mode ($2\nu\beta\beta$) is allowed in the Standard Model and has been observed in more than 10 nuclei \cite{Barabash:2015eza}. In this case, the summed energy of the two electrons is a continuum, while for the $0\nu\beta\beta$-decay the distinct signature is a peak at the Q-value, the mass difference between the mother and daughter nuclei.

Experiments can observe a certain decay rate in a detector. The corresponding half-life is inversely proportional to $\langle m_{\beta\beta}\rangle^2$,

\begin{equation}
\frac{1}{T_{1/2}^{0\nu}} = \frac{\langle m_{\beta\beta}\rangle^2}{m_e^2} G^{0\nu} |M^{0\nu}|^2,
\label{eq:Majorana_mass}
\end{equation}

\noindent
assuming that the decay is mediated by the exchange of a light Majorana neutrino. $m_e$ is the mass of the electron, $G^{0\nu}$ is the phase space factor, and $M^{0\nu}$ is the nuclear matrix element. Recent experimental limits on $T_{1/2}^{0\nu}$ and $\langle m_{\beta\beta}\rangle$ are of the order $T_{1/2}^{0\nu} \ge ($10$^{25}$-10$^{26}$)\,yr and $\langle m_{\beta\beta}\rangle \le (0.06-0.17)$\,eV, using a variety of nuclei and detector technologies~\cite{Henning:2016fad,Dolinski:2019nrj}.  

A particularly suitable isotope to search for the $0\nu\beta\beta$-decay with is $^{136}$Xe, with  $Q_{\beta\beta}$=(2457.83$\pm$0.37)\,keV~\cite{Redshaw:2007un}. Current experiments use liquid xenon either in its pure form, EXO-200~\cite{Anton:2019wmi}, or {\color{black}xenon dissolved in liquid scintillator}, KamLAND-Zen~\cite{KamLAND-Zen:2016pfg}, and provide competitive constraints on the half-life. Future detectors that use xenon gas operated at high pressure, NEXT~\cite{Alvarez:2012flf,Gomez_NEXT:2019} and PandaX-III~\cite{2017SCPMA..60f1011C}, will add tracking capabilities for improved background rejection, while nEXO~\cite{Albert:2017hjq} proposes to operate a total of 5\,t of isotopically enriched liquid xenon.

DARWIN~\cite{Aalbers:2016jon} is a proposed observatory using \SI{40}{t} of liquid natural xenon (LXe) in a time projection chamber (TPC) with the primary goal of searching for particle dark matter. Here, we demonstrate that DARWIN has a similar reach to dedicated future neutrinoless double beta decay experiments. This is due to its large, homogeneous target, and its ultra-low background, coupled to the capability of the TPC to simultaneously measure the location, energy, particle type and multiplicity of an event \cite{Schumann:2014uva}.

The paper is organized as follows: in Sect.~\ref{sec:darwin} we provide a brief review of the baseline design of the DARWIN detector and describe the detector model utilized in our simulation study. Sect.~\ref{sec:signal} addresses the signal topology and how it is used to reject background events.
In Sect.~\ref{sec:bg} we discuss the expected background sources, while the resulting background spectra and rates are presented in Sect.~\ref{sec:analysis}. We discuss DARWIN's sensitivity to $0\nu\beta\beta$-decay in Sect.~\ref{sec:sensitivity} and give a summary and an outlook in Sect.~\ref{sec:summary}.

\section{The DARWIN Observatory}
\label{sec:darwin}

DARWIN is a next-generation dark matter experiment that will operate a \SI{40}{t} active (\SI{50}{t} total) liquid xenon TPC with the main goal to probe the entire experimentally accessible parameter space for weakly interacting massive particles (WIMPs) as dark matter candidates.  
Other physics goals include the search for the $0\nu\beta\beta$-decay, the real-time detection of solar $pp$ neutrinos via electron scattering, the observation of supernova and solar $^8$B neutrinos via coherent neutrino nucleus scattering and the search for solar axions, galactic axion-like particles and dark photons. 

The DARWIN detector is described in detail in~\cite{Aalbers:2016jon}.  In the baseline scenario, the detector is a cylindrical, two-phase (liquid and gas) xenon TPC with  \SI{2.6}{m} diameter and \SI{2.6}{m} height. The TPC will be placed in a low-background, double-walled cryostat surrounded by an instrumented water tank to shield it from the environmental radioactivity and to record the passage of cosmic muons and their secondaries as well as for neutron thermalization. 

Interactions in the TPC will give rise to a prompt signal ($S1$) from photons and a delayed, proportional scintillation signal ($S2$) from electrons transported by a homogeneous drift field and extracted into the gas phase. Both signals will be detected by photosensor arrays (made of photomultiplier tubes (PMTs),  silicon photomultiplier (SiPM), or new types of sensors), providing the $x$-$y$-$z$-coordinates of an interaction, as well as its energy with $<$\,1\% $1\,\sigma$ resolution for MeV energy depositions. 
Interactions separated by more than \SI{15}{mm} are assumed to be individually identified in event reconstruction.
This allows for separation between single scatters (as expected from $0\nu\beta\beta$-decays and dark matter particle interactions) and multiple scatters (as expected from many sources of backgrounds), as well as the definition of an inner (fiducial) volume with reduced background levels. The high density of the liquid xenon ($\sim$3\,g/cm$^3$) ensures a short attenuation length for $\gamma$-rays.

The final location of the DARWIN experiment is yet to be decided. A good candidate is the Gran Sasso Underground Laboratory (LNGS) in Italy. We will use its overburden in this study. 

\subsection{Monte Carlo model of the detector}
\label{sec:darwin_MCmodel}

For the Monte Carlo event generation and particle propagation in {\sc{geant4}} we use a realistic model of the DARWIN detector. Its details are described in the following. 

\begin{figure}[ht!]
\centering
\includegraphics[width=\linewidth]{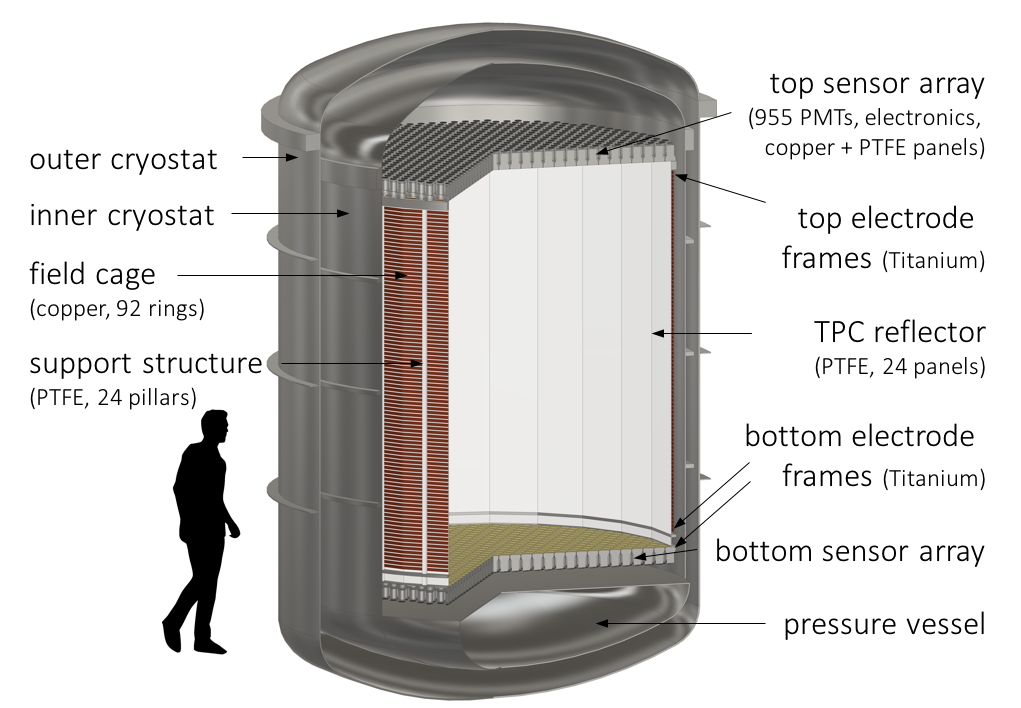}
\caption{Drawing of DARWIN's double-walled cryostat and TPC, showing all components considered in the simulation.}
\label{fig:cryostat}
\end{figure}

The TPC is enclosed within the outer and  inner titanium cryostat (shown in Fig.~\ref{fig:cryostat}), including torispherical domes, flanges and stiffening rings to minimize the amount of material. A {\color{black}dome-shaped} pressurizable titanium vessel is placed on the inner cryostat floor to reduce the {\color{black}volume to be filled with liquid} xenon while keeping the material budget low.
A study based on previously-measured specific activities of cryostat materials \cite{Akerib:2017iwt,Aprile:2017ilq} showed that a cryostat made of titanium yields a lower background rate than a stainless steel cryostat of equal mechanical {\color{black}properties}. 

The inner cryostat contains the  liquid xenon volume and the TPC. The TPC walls are formed by PTFE reflectors of 3 mm thickness with high reflectivity for the vacuum ultra-violet (VUV) scintillation light, surrounded by 92 cylindrical copper field shaping rings. The structure is reinforced with 24 PTFE support pillars. Titanium frames at the bottom and top of the TPC support the electrodes to establish drift and extraction fields. Two photosensor arrays are located at the top and bottom of the TPC cylinder, consisting of a structural copper support, a PTFE reflector disk, the VUV-sensitive photosensors and the sensors' cold electronics. Because the final sensor type is yet to be chosen for DARWIN and R\&D on light sensor options~\cite{Baudis:2018pdv,Baudis:2013xva,Arneodo:2018sip,Ferenc:2017aba} is ongoing, the top and bottom sensors have, for the majority of simulations, been simplified to two disks which properly account for the material budget and the associated activities of radioactive isotopes. This allows for a direct comparison between a baseline scenario with PMTs and an alternative based on SiPMs. 

All the major components included in the simulations are listed in Table~\ref{tab:geometryelements}. The assumed radioactivity levels of the materials are discussed in Sect. \ref{sec:bg} and listed in Table \ref{tab:materialcontamination}.

\begin{table}[ht]
\begin{center}
\begin{tabular}{lcc}
\hline
    Component                             & Material  & Mass   \\
\hline
    Outer cryostat                        & Titanium  &  \SI{3.0}{t}   \\
    Inner cryostat                        & Titanium  &  \SI{2.1}{t}   \\
    Bottom pressure vessel                & Titanium  &  \SI{0.4}{t}   \\
\hline    
    LXe instrumented target               & LXe	      &  \SI{39.3}{t}   \\
    LXe buffer outside the TPC            & LXe		  &  \SI{9.0}{t}   \\
    LXe around pressure vessel            & LXe       &  \SI{270}{kg}   \\
    GXe in top dome + TPC top             & GXe       &  \SI{30}{kg}    \\
\hline
    TPC reflector (3mm thickness)         & PTFE      &  \SI{146}{kg}   \\
    Structural support pillars (24 units) & PTFE      &  \SI{84}{kg}    \\
    Electrode frames                      & Titanium  &  \SI{120}{kg}   \\
    Field shaping rings (92 units)        & Copper    &  \SI{680}{kg}   \\
\hline
    Photosensor arrays (2 disks):         &           &                 \\
    Disk structural support               & Copper    &  \SI{520}{kg}   \\
    Reflector + sliding panels            & PTFE      &  \SI{70}{kg}    \\
    Photosensors: 3"~PMTs (1910 units)    & composite &  \SI{363}{kg}   \\ 
    Sensor electronics (1910 units)       & composite & \SI{5.7}{kg}    \\
\hline
\end{tabular}
\caption{List of detector components included in the {\sc{geant4}} geometry model of DARWIN stating their material composition and total mass.}
\label{tab:geometryelements}
\end{center} 
\end{table}

\section{\texorpdfstring{$0\nu\beta\beta$}{} signal events in liquid xenon}
\label{sec:signal}

\begin{figure}[ht!]
  \centering
  \includegraphics[width=1.\linewidth]{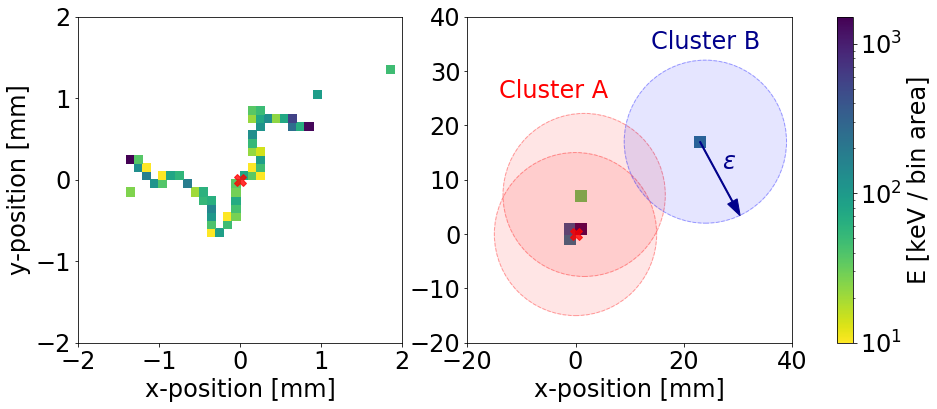}
\caption{Simulated energy deposition (color scale) of two different $0\nu\beta\beta$-events in the $x$-$y$-plane. Left: The two electrons thermalize along two {\color{black}non-resolvable} back-to-back tracks. The emitted Bremsstrahlung photons yield detached energy depositions. Right: A $\mathcal{O}$(\SI{400}{keV}) photon Compton scatters \SI{8}{mm} from the position of the decaying $^{136}$Xe nucleus and travels more than \SI{2}{cm} without energy loss before absorption. The circles indicate {\color{black}the boundaries of individually resolvable clusters assuming } a separation threshold $\epsilon  = \SI{15}{mm}$.}
\label{fig:topology}
\end{figure}

In a $0\nu\beta\beta$-decay, the energy $Q_{\beta\beta}$ is released mainly in the form of kinetic energy of the two electrons. In liquid xenon, the electrons thermalize within $\mathcal{O}$(mm) resulting in a single site (SS) signal topology, as shown in Fig.~\ref{fig:topology}~(left).
Bremsstrahlung photons emitted during electron thermalization travel some distance without energy deposition before scattering or being absorbed. 
Abundantly emitted low energy photons are likely to deposit their energy close to the decay position and remain unresolved in the DARWIN detector. Photons with energies above $\SI{300}{keV}$ have a mean free path of more than \SI{15}{mm} and might travel larger distances before interacting. This can result in an energy deposition which is spatially separable and can cause a false identification as a multi site (MS) event, Fig.~\ref{fig:topology}~(right). 

\begin{figure}[ht!]
  \centering
  \includegraphics[width=1.\linewidth]{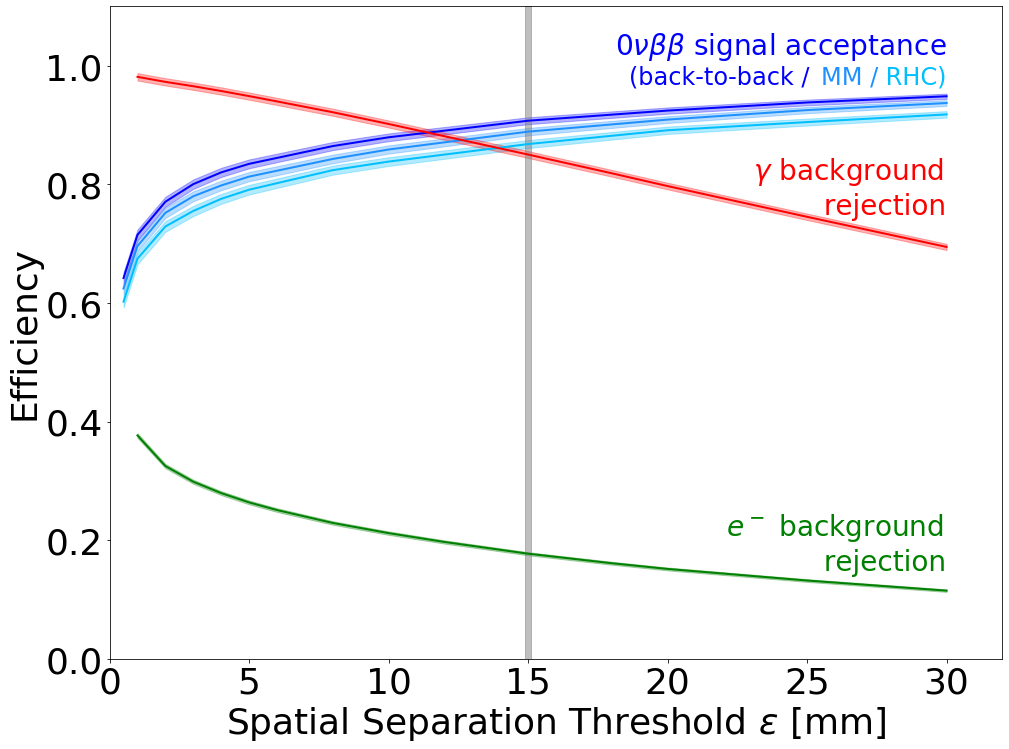}
\caption{ {\color{black}Efficiency of $0\nu\beta\beta$ signal acceptance and background rejection as a function of the spatial separation threshold $\epsilon$. The three signal lines (blue) compare different energy and angular distributions based on a back-to-back electron emission, a mass mixing (MM) mechanism and a right-handed current (RHC) model. The background rejection efficiency is shown for $\gamma$s (red) and electrons (green) with $E = Q_{\beta\beta}$. The vertical line (grey) corresponds to the value of $\epsilon$ assumed in this study. Bands indicate $\pm 2\,\sigma$ uncertainties.} }
\label{fig:Acceptance}
\end{figure}

Energy depositions are therefore spatially grouped using a density-based spatial clustering algorithm~\cite{Ester96adensity-based}. An energy deposition is considered as a new cluster if its distance to any previous energy deposition is larger than our selected separation threshold~$\epsilon$.
{\color{black}Fig.~\ref{fig:Acceptance} shows the efficiency for signal acceptance and background rejection for photons and electrons with an energy of $Q_{\beta\beta}$ as a function of $\epsilon$.}

The distribution of energy per electron and the angle between the two depend on the yet unknown decay mechanism. We assume a mass mixing mechanism and the most probable decay where the electrons are emitted back-to-back, each with a kinetic energy of $Q_{\beta\beta}/2$. This assumption is compared {\color{black}in Fig.~\ref{fig:Acceptance}} to the predicted energy and angular distributions in the mass mixing (MM) model and a right-handed current (RHC) model presented in~\cite{Arnold_2010}. 

We assume that a spatial separation between energy depositions of $\epsilon = \SI{15}{mm}$ can be resolved in the DARWIN TPC. This results in a signal acceptance of $90.4\%$ (MM: $88.7\%$, RHC: $86.6\%$) as SS events. Background events from electrons and photons with $Q_{\beta\beta}$ energy are rejected  as MS with an efficiency of $17.7\%$ and $85.1\%$, respectively. 
A smaller separation threshold  $\epsilon$ results in a larger fraction of misidentified $0\nu\beta\beta$-decays. Simultaneously, electrons from $\beta$ decays and $\gamma$-ray events are more efficiently identified as MS. 
As we will discuss in Sect.~\ref{sec:summary}, a lower spatial threshold can increase the sensitivity to $0\nu\beta\beta$-decays. The decrease in signal acceptance is overcompensated by the improved background rejection. 

\section{Background events in the \texorpdfstring{$0\nu\beta\beta$}{} energy range}
\label{sec:bg}

We discuss all background sources which contribute events within the energy range of [2.3 - 2.7]~MeV around $Q_{\beta\beta}$.
We consider intrinsic background events from radioactive decays (radiogenic) and those induced by cosmic neutrinos, muons and their secondaries (cosmogenic).
Intrinsic events are homogeneously distributed in the liquid xenon. 
Likewise we study radiogenic background radiation from external sources emanating into the target.
 
\begin{table*}[ht]
\begin{center}
\begin{tabular}{lcccccccc}
\hline
  Material & Unit    & $^{238}$U& $^{226}$Ra & $^{232}$Th & $^{228}$Th  & $^{60}$Co  &  $^{44}$Ti & Reference \\
\hline
Titanium & mBq/kg    &   $<$1.6   & $<$0.09  &   0.28    &  0.25    &   $<$0.02    &  $<$1.16 &~\cite{Akerib:2017iwt} \\
PTFE  & mBq/kg       &   $<$1.2   &   0.07   &  $<$0.07  &  0.06    &  0.027       &    - &~\cite{Aprile:2017ilq}  \\
Copper  & mBq/kg     &   $<$1.0   & $<$0.035 &  $<$0.033  &  $<$0.026 &   $<$0.019      &  - &~\cite{Aprile:2017ilq} \\
PMT & mBq/unit       &   8.0    & 0.6      &   0.7      &  0.6      &    0.84       & -  &~\cite{Aprile:2017ilq} \\
Electronics  & mBq/unit   &  1.10    & 0.34     &   0.16     &  0.16     &    $<$0.008  & - & ~\cite{Aprile:2017ilq}\\
\hline
\end{tabular}
\caption{Assumed activity levels for the simulated materials and isotopes.}
\label{tab:materialcontamination}
\end{center} 
\vspace{-0.2cm}
\end{table*}
 
\subsection{Homogeneously distributed intrinsic background}
\label{sec:bg_intrinsic}

The intrinsic background sources originate from noble gas isotopes or from interactions of cosmogenic particles with the xenon target: 

\begin{itemize}

\item $^{8}$B solar neutrinos are an irreducible background source. The expected rate of $\nu$-$e^-$ scatterings is derived assuming a $^{8}$B-$\nu$ flux of $\phi = (5.46 \pm 0.66)\times10^{6}$~cm$^{-2}$s$^{-1}$~\cite{Tanabashi:2018oca}. The calculation of scattering cross sections follows~\cite{dutta2019neutrino}. The electron neutrino survival probability is conservatively estimated to be $P_{ee} = 0.50^{+5 \%}_{-30 \%}$ for neutrinos with $E_\nu > Q_{\beta\beta}$.

\item $^{137}$Xe from cosmogenic activation: muon-induced neutrons produced in the liquid xenon can thermalize and be captured on a $^{136}$Xe nucleus, producing $^{137}$Xe, {\color{black} as measured by EXO-200~\cite{EXO200::2015wtc}}. This isotope decays via a $\beta^{-}$ process with Q$_\beta$ = 4.17\,MeV and a half-life of 3.82\,min. Assuming the depth of LNGS and previous simulations of the muon-induced neutron flux underground~\cite{Selvi:2011zz}, we estimate the muon-induced $^{137}$Xe production rate in DARWIN to be $(6.9 \pm 0.4)$ atoms/(t$\cdot$yr). Neutrons  produced in the solid materials contribute about 5\% of this rate.
Activation of $^{136}$Xe due to radiogenic neutrons from the TPC  materials has been found to be subdominant by more than two orders of magnitude. Activation of xenon in the non-shielded environment of the purification loop is non-negligible, but can be efficiently suppressed by a delayed re-feed of the LXe into the detector. {\color{black} Suppression by three orders of magnitude adds an additional \SI{225}{kg} to the total xenon budget when cycling \SI{1000}{} standard liter per minute}. 

\item The $2\nu\beta\beta$ decay spectrum of $^{136}$Xe has been simulated assuming the measured half-life of $T_{1/2}=(2.165\pm0.061)\times10^{21}$\,yr~\cite{Albert:2013gpz}. For the analytic spectrum we use the non-relativistic Primakoff-Rosen approximation for the interaction between nuclei and electrons in the parametrization discussed in~\cite{boehm_vogel_1992}. This approximation is conservative as it overestimates the rate around the spectral end point. 

\item $^{222}$Rn in LXe is assumed to be reduced by online cryogenic distillation~\cite{Aprile:2017kop} and stringent material selection to a concentration equivalent to 0.1\,$\mu$Bq $^{222}$Rn activity per kg of xenon. Being crucial for the WIMP search, significant efforts are being undertaken to reach this design goal. The dominant intrinsic background contribution for the $0\nu\beta\beta$ search originates from the $\beta$-decay of $^{214}$Bi (Q$_\beta$ = 3.27\,MeV). In 19.1$\%$ of the cases it decays to the $^{214}$Po ground state without $\gamma$-emission, which renders the rejection based on spatial topology rather inefficient, as discussed in Sect.~\ref{sec:signal}. The short half-life of the decay daughter $^{214}$Po ($T_{1/2} = \SI{164.3}{\mu s}$), however, allows for BiPo event tagging and suppression with more than \SI{99.8}{\%} efficiency~\cite{Baudis:2013qla}.

\end{itemize}

\subsection{External radiogenic background sources} 
\label{sec:bg_external}

Long-lived radionuclides are present in each detector material. Their decays, as well as the subsequent decays of their daughter isotopes, might introduce background in the target. 
Activity levels for all materials are listed in Table~\ref{tab:materialcontamination} and based on 
reports from previous or ongoing experiments~\cite{Akerib:2017iwt,Aprile:2017ilq}.

\begin{itemize}

\item The natural decay chains of $^{238}$U, $^{232}$Th and $^{235}$U yield a background contribution primarily from $\gamma$-rays emitted by $^{214}$Bi- (E$_\gamma$ = 2.45\,MeV) and $^{208}$Tl-decays (E$_\gamma$ = 2.61\,MeV). The former two chains were split into their early and late component at $^{226}$Ra and $^{228}$Th, respectively, to account for radiogenic non-equilibrium.  

\item $^{60}$Co $\beta$-decays dominantly (99.95\%) via the two excited states of $^{60}$Ni. The de-excitation is temporally non-resolvable and spatial coincidences of the 1.17\,MeV and the 1.33\,MeV $\gamma$-events contribute to the background. 

\item Among the radio-isotopes from cosmogenic material activation at sea level~\cite{Zhang:2016rlz}, $^{44}$Ti in the cryostat material is the most relevant, due to its long half-life ($T_{1/2}$ = 59.1\,yr) and the subsequent decay of $^{44}$Sc with $\gamma$-emission at 2.66\,MeV. 

\item $^{222}$Rn contamination in the non-instrumented xenon surrounding the TPC can contribute to the $^{214}$Bi-induced $\gamma$-background. The rejection based on BiPo tagging described above cannot be applied since the subsequent alpha decays are not observed. 

\end{itemize}

\section{Analysis and background results}
\label{sec:analysis}

The background sources discussed in Sect.~\ref{sec:bg} are simulated with the {\sc{geant4}} particle physics simulation toolkit~\cite{Agostinelli:2002hh}, using the detector model presented in Sect.~\ref{sec:darwin_MCmodel}. The equivalent of at least 100 years of DARWIN run time has been simulated for each material and isotope. In this section, we discuss the methods applied for event selection. The analytical background model, used for the profile-likelihood analysis in Sect.~\ref{sec:Sensitivity_Freq}, is also described, and the background results are discussed. 

\subsection{Monte Carlo data processing and event selection}
\label{sec:analysis_DataProc}

The energy depositions generated by {\sc{geant4}} per event undergo a density-based spatial clustering algorithm~\cite{Ester96adensity-based} to topologically distinguish signal-like single site (SS) from background-like multi site (MS) events, as discussed in Sect.~\ref{sec:signal}. We assume a separation threshold $\epsilon = \SI{15}{mm}$ for the DARWIN TPC. This comparatively coarse clustering inevitably results in a fraction of $\gamma$-accompanied $\beta$-decays from background events, e.g., $^{214}$Bi decays which frequently occur with higher multiplicity, being falsely identified as SS and consequentially contributing to the background. 

To account for the finite energy resolution of the detector, the combined energy deposited inside each cluster is smeared according to a resolution of
\begin{equation}
\frac{\sigma_E}{E} = \frac{a}{\sqrt{E[\mathrm{keV}]}} + b,
\label{eq:E_resolution}
\end{equation}
\noindent
with $ a = (0.3171 \pm 0.0065)$  and  $b=(0.0015 \pm 0.0002)$. At $E = Q_{\beta\beta}$ this corresponds to $\sigma_E / E= 0.8 \%$, as demonstrated in the XENON1T TPC~\cite{highE-rec}.
The cluster position is smeared to account for the detector's spatial resolution which is conservatively assumed to be $\sigma_{x,y}= \sigma_{z} = \SI{10}{mm}$ above $\SI{2}{MeV}$. 

\begin{figure}[!t]
\begin{center}
\includegraphics[width =\linewidth]{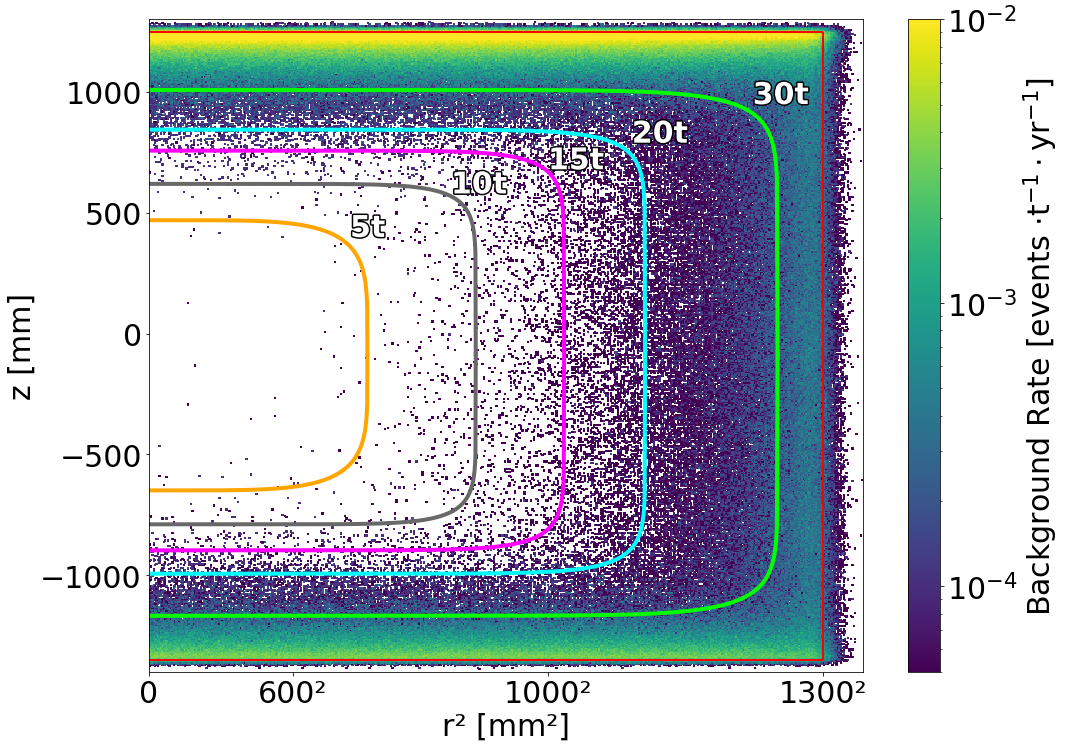}
\caption{Spatial distribution of external background events inside the instrumented volume for 100 years of DARWIN run time. The colored lines indicate the contours of the optimized fiducial volumes containing different LXe target masses. The \SI{5}{t} fiducial volume is used for the sensitivity estimate presented below.}
\label{fig:fv}
\end{center}
\vspace{-5mm}
\end{figure}

Constraining the target to a super-ellipsoidal-shaped fiducial volume (FV) allows us to exploit the excellent self-shielding capabilities of liquid xenon. To compensate for the reduced shielding power in the xenon gas phase, the FV is shifted slightly downwards from the center of the instrumented volume. 
The fiducial volume is optimized for each FV mass independently. We use the lifetime-weighted combined external background, after the selection of single site events, energy and spatial resolution smearing. 
Only events with an energy inside the $0\nu\beta\beta$-ROI of [2435 - 2481]~keV, defined as the full width at half maximum (FWHM) range of the expected signal peak, are considered.  
The spatial distribution of external background events inside the active volume is shown in Fig. \ref{fig:fv}.

\subsection{Background model and fiducial mass dependence}
\label{sec:analysis_BackgroundResults_Model}

\begin{figure}[!ht]
\includegraphics[width=0.97\linewidth]{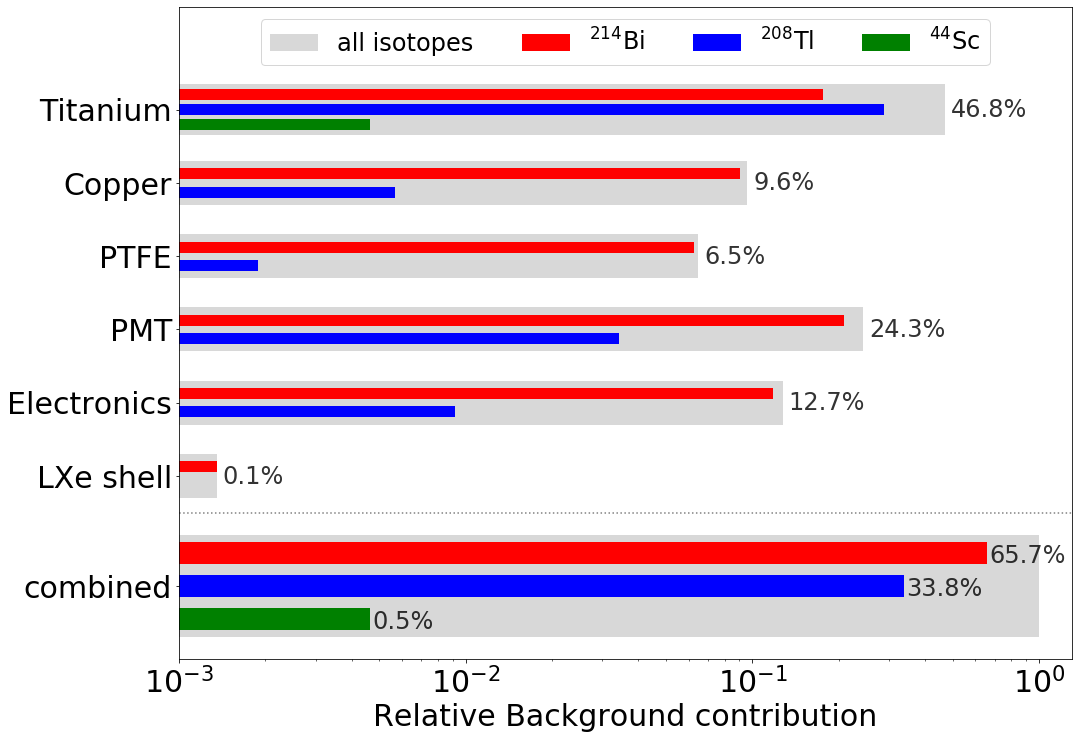}
\begin{center}
\includegraphics[width = \linewidth]{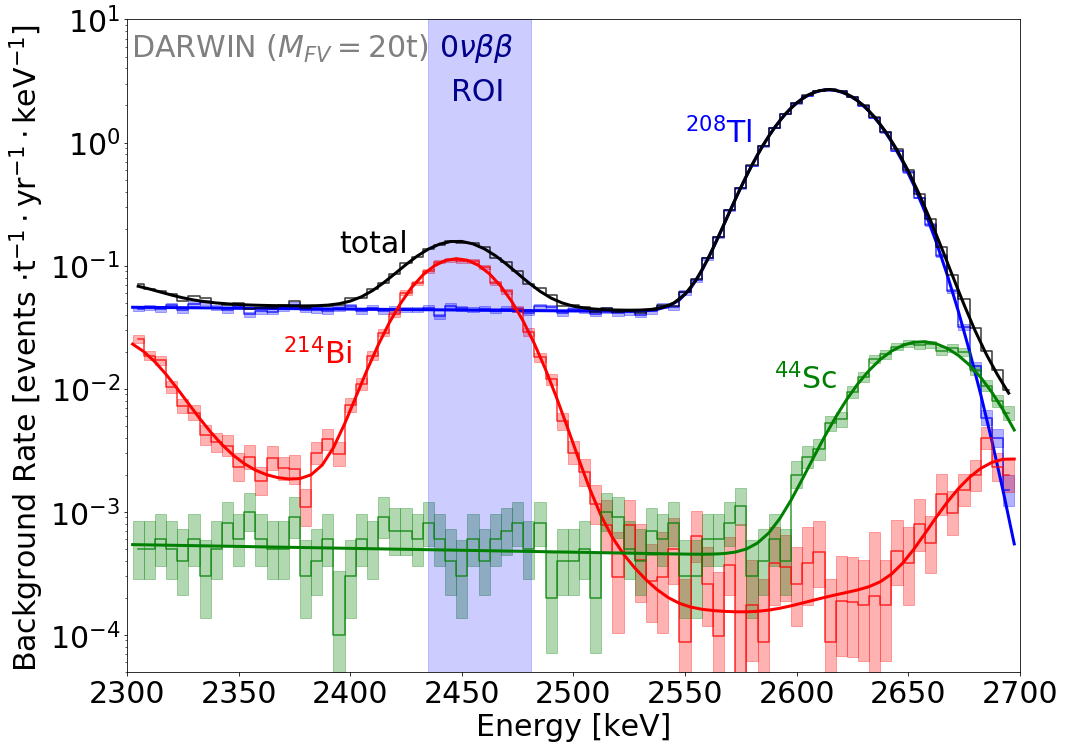}
\caption{Composition of the material-induced external background in the \SI{20}{t} fiducial volume. Top: Relative contribution to the background in the $0\nu\beta\beta$-ROI by material and isotope. Bottom: Background spectra by isotope with the corresponding model fits. The relative contributions and spectral shapes are representative for smaller fiducial volumes.}
\label{fig:materialcontributions}
\end{center}
\vspace{-5mm}
\end{figure}

The selection of events within a fiducial volume removes all $\alpha$- and $\beta$-contributions originating from external sources. The $\gamma$-background is shown in Fig. \ref{fig:materialcontributions} (bottom) for the \SI{20}{t} fiducial volume. In the $0\nu\beta\beta$-ROI the background is composed of the absorption peak from $^{214}$Bi at $E_{\mathrm{Bi}} = \SI{2.45}{MeV}$ and Compton scattered photons, mainly from the $^{208}$Tl line ($E_{\mathrm{Tl}} = \SI{2.61}{MeV}$). Compton scatterings inside the fiducial volume with the subsequent escape of the scattered lower energy $\gamma$-ray are strongly suppressed by fiducialization. The continuous background is dominated by photons that undergo an undetected Compton scatter outside the detector followed by their absorption in the fiducial volume. 

The continuous contribution from $\gamma$-rays emitted in $^{44}$Sc decays accounts for less than 1\% of the external background. 
$^{214}$Bi decays with $E_\gamma > Q_{\beta\beta}$ contribute with a similarly subdominant level. 
Spatial coincident absorption of both $^{60}$Co gammas accounts for only approximately $10^{-3}$ of the total material background at $E = \SI{2.51}{MeV}$ in the \SI{30}{t} fiducial volume. 
In the fiducial volume mass range of interest, it can be considered negligible. 
The largest background contribution in the ROI is induced by the absorption peak of \SI{2.45}{MeV} $\gamma$-rays emitted by $^{214}$Bi decays in the detector materials. 
The contribution from $^{214}$Bi decays in the non-instrumented LXe around the TPC accounts for approximately 0.1\% of the total material-induced background. 

The relative contributions to the $\gamma$-background in the ROI are shown per material of origin in Fig. \ref{fig:materialcontributions}~(top). The similar contribution of cryostat-induced events from the walls and the combined PMT and electronics background originating from the top and bottom sensor array is a result of the optimization of the fiducial volume, which is properly balancing the $r$- and $z$-extent. 

The spectral shape of the material-induced $\gamma$-back-ground is modelled with a Gaussian peak and an exponentially decreasing continuum for each line, as shown in Fig. \ref{fig:materialcontributions}~(bottom). We consider the \SI{2.61}{MeV} $^{208}$Tl peak, the \SI{2.66}{MeV} $^{44}$Sc peak and each contribution of $^{214}$Bi with $E_\gamma > \SI{2.0}{MeV}$. The ratio between the $^{214}$Bi and the $^{44}$Sc peaks to the $^{208}$Tl peak intensity is established using Monte Carlo data in fiducial volumes sufficiently large to provide high statistics. Similarly, each continuum contribution is tied to its corresponding peak intensity and a fixed relation between the three slope parameters is found. 
The only remaining free parameters of the combined model are the total intensity of the $^{208}$Tl peak and one common slope parameter. The model is tested and confirmed using a $\chi^2$ goodness-of-fit test on the combined external background in the fiducial mass range $\leq \SI{20}{t}$. 

The intrinsic background from $^8$B neutrinos is assumed to be flat. The spectra corresponding to $^{137}$Xe and $^{222}$Rn are approximated linearly falling in the [2.2-\SI{2.8}{MeV}] range. The slopes are obtained from Monte Carlo studies. The $2\nu\beta\beta$ spectrum is convolved with the Gaussian energy resolution. 

\begin{figure}[ht]
\centering
\includegraphics[width = \linewidth]{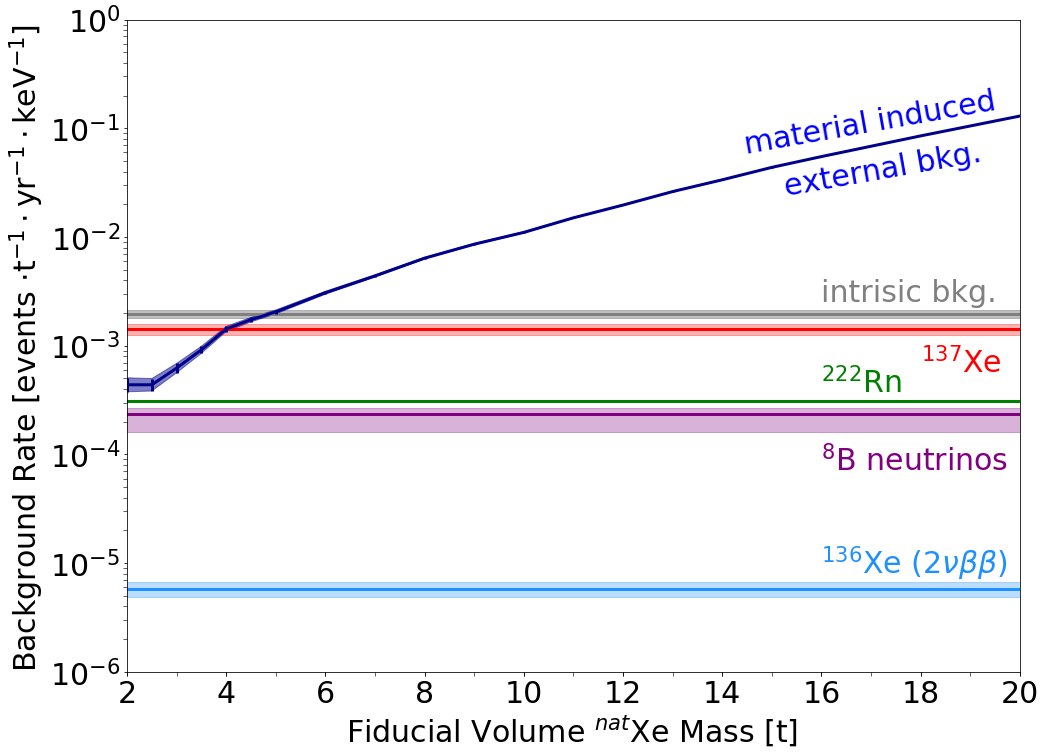}
\caption{Background rate in the ROI versus fiducial mass. External contributions are combined. Fiducial volume independent intrinsic sources are shown per contribution. Bands indicate $\pm 1\,\sigma$ uncertainties. At $\SI{5}{t}$, the external sources contribute at the same level as the combined intrinsic background.}
\label{fig:bg_mass}
\end{figure}

The suppression of the external background with decreasing fiducial mass is shown in Fig. \ref{fig:bg_mass}, together with the target mass
 independent intrinsic contributions. 

\subsection{Background rates in the $0\nu\beta\beta$-ROI}
\label{sec:analysis_BackgroundResults_ROI}

\begin{figure}[ht]
\begin{center}
\includegraphics[width=\linewidth]{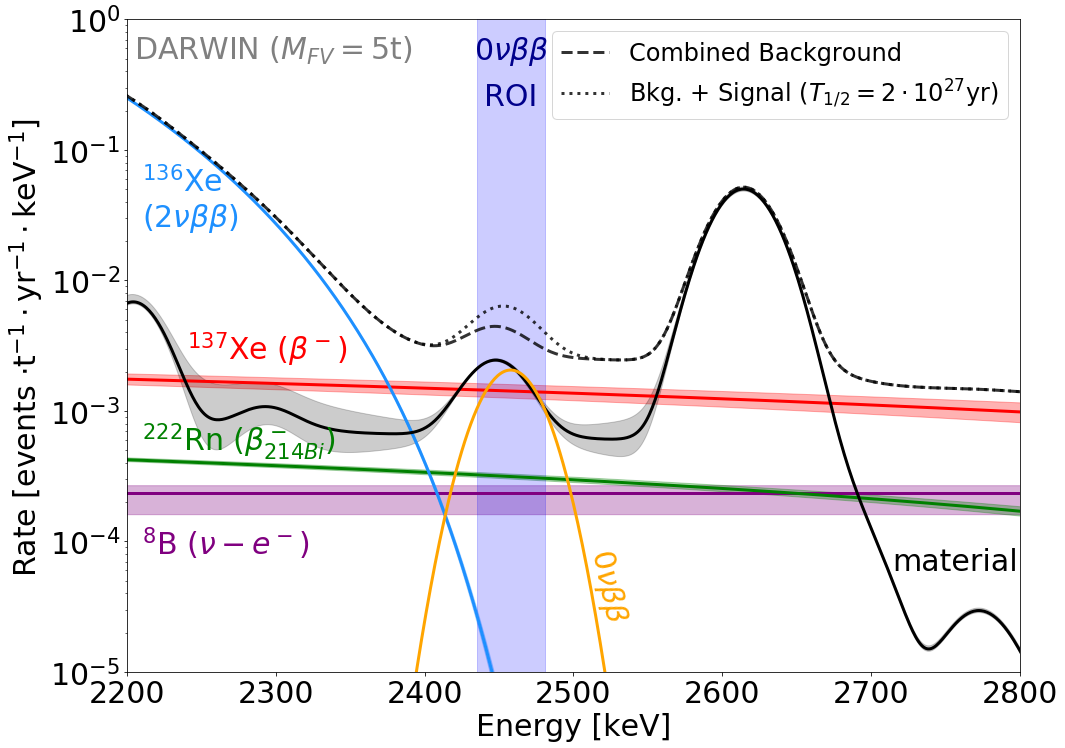}
\caption{Predicted background spectrum around the $0\nu\beta\beta$-ROI for the \SI{5}{t} fiducial volume. A hypothetical signal of 0.5 counts per year corresponding to $T^{0\nu}_{1/2} \approx \SI{2e27}{yr}$ is shown for comparison. Bands indicate $\pm 1\,\sigma$ uncertainties.}
\label{fig:background_spectrum_material} 
\end{center}
\end{figure}

The fiducial volume is optimized for $T^{0\nu}_{1/2}$ sensitivity, as discussed in detail in Sect.~\ref{sec:Sensitivity_F-o-M}, and yields \SI{5}{t}. The resulting background spectrum from intrinsic and external sources is shown for this fiducial mass in Fig.~\ref{fig:background_spectrum_material}.

\begin{table*}[!ht]
\begin{center}
\begin{tabular}{l c c c}
\hline
Background source  & Background index &  Rate & Rel. uncertainty \\
                    & [events/(t$\cdot$yr$\cdot$keV )] & [events/yr]  & \\
\hline
\multicolumn{3}{l}{{\it External sources ({\color{black} \SI{5}{t}} FV):}}  \\
  $^{214}$Bi peaks + continuum  &  $1.36 \times10^{-3}$  & 0.313 & $\pm 3.6\%$\\
  $^{208}$Tl continuum       &  $6.20 \times10^{-4}$ & 0.143 & $\pm 4.9\%$ \\
  $^{44}$Sc continuum        &  $4.64  \times10^{-6}$ & 0.001 & $\pm 15.8\%$ \\
\hline
\multicolumn{3}{l}{{\it Intrinsic contributions:}} \\
  $^{8}$B ($\nu-e$ scattering)           &  $2.36 \times10^{-4}$ & 0.054 & $ +13.9\%, -32.2\%$\\
  $^{137}$Xe ($\mu$-induced $n$-capture) &  $1.42 \times10^{-3}$  & 0.327 & $\pm 12.0\%$ \\
  $^{136}$Xe $2\nu\beta\beta$            &  $5.78 \times10^{-6}$  & 0.001 & $+17.0\%, -15.2\%$\\   
  $^{222}$Rn in LXe (0.1\,$\mu$Bq/kg)    &  $3.09 \times10^{-4}$ & 0.071 &  $\pm 1.6\%$ \\
\hline
  \textbf{Total:}                        & $\mathbf{3.96\times 10^{-3}}$ & $\mathbf{0.910}$ & $\mathbf{ +4.7\%, -5.0\%}$\\
\hline
\end{tabular}
\captionsetup{width=.75\textwidth}
\caption{Expected background index averaged in the $0\nu\beta\beta$-ROI of [2435 - 2481]~keV, corresponding event rate in the \SI{5}{t} FV and relative uncertainty by origin.}
\label{tab:backgroundsources}
\end{center} 
\end{table*}

The intrinsic background in the ROI is dominated by the gently falling $\beta^-$-spectrum of $^{137}$Xe decay. Subdominant contributions are the electron scattering of solar $^8$B neutrinos and $\beta^-$-events from $^{214}$Bi-decays which are not vetoed by BiPo tagging. The $2\nu\beta\beta$ spectrum overlaps negligibly with the ROI, but dominates the background toward lower energies. 

The model-estimated background indices for all contributions are summarized in Table~\ref{tab:backgroundsources}. 
To validate the analytic model introduced in Sect.~\ref{sec:analysis_BackgroundResults_Model}, we compare the background model estimate with the values derived by weighted event counting in the \SI{5}{t} fiducial mass data from Monte Carlo. Both results are in agreement within the statistical errors. The model-derived uncertainty on the background, however, is a factor of~4 lower than the Poissonian statistics error in the simple counting approach. 
The uncertainties on intrinsic background sources account for statistical errors, the variation of the overlap with the $0\nu\beta\beta$-ROI based on the energy resolution and systematic uncertainties from (theory-driven) input parameters. The dominant contributions are the $\nu_e$ survival probability and the neutrino flux ($^8$B $\nu$-$e^-$ scattering), the $^{136}$Xe neutron capture cross-section (governing the $^{137}$Xe production rate) and the half-life of $^{136}$Xe ($2\nu\beta\beta$ decay).

\section{Sensitivity Calculation}
\label{sec:sensitivity}
We use the background rates predicted in Sect.~\ref{sec:analysis_BackgroundResults_ROI} to derive a limit on the half-life sensitivity at 90\% confidence level (C.L.) as well as the $3\,\sigma$ discovery potential for the $0\nu\beta\beta$-decay. The latter is defined as the minimal value of $T_{1/2}^{0\nu}$ required to exclude the null hypothesis with a median significance of 99.7\%~C.L. 

\subsection{Half-life sensitivity estimation}
\label{sec:Sensitivity_F-o-M}
Based on the figure-of-merit estimator proposed in~\cite{Avignone:2005cs} we calculate the half-life sensitivity at 90\%~C.L. as:
\begin{equation}
     T_{1/2}^{0\nu} =  \ln{2} \frac{\epsilon \: f_{\textrm{ROI}}\: \alpha \: N_{A}}{1.64 \: M_{\mathrm{Xe}}}  \frac{\sqrt{Mt}}{\sqrt{B\Delta E}} ,
\label{eq:figure_merit}
\end{equation}
\noindent
with $\epsilon = 0.9$ being the detection efficiency of a single site $0\nu\beta\beta$-decay event, $f_{\textrm{ROI}}=0.76$ the fraction of signal covered by the ROI,  $\alpha = 0.089 $ the abundance of $^{136}$Xe in natural xenon, $N_{A}$ the Avogadro number in mol$^{-1}$, $M_{\textit{Xe}}$ the molar mass number of xenon in t/mol, $M$ the fiducial mass in tons, $t$ the exposure time in years, $B$ the background index in t$^{-1}$yr$^{-1}$keV$^{-1}$, and $\Delta E$ the width of the ROI in keV. The value 1.64 is the number of standard deviations corresponding to a 90\%~C.L.

Following Eq.~(\ref{eq:figure_merit}) and using the background index for the \SI{5}{t} fiducial mass (Table~\ref{tab:backgroundsources}), we obtain a half-life sensitivity of {\color{black}\SI{2.0e27}{yr} } ({\color{black}\SI{1.3e27}{yr}}) after 10 (4) years of exposure. 

This figure-of-merit estimation is an established tool to directly compare $0\nu\beta\beta$ sensitivities of different experiments using common statistical methods and assumptions. It also allows for a straightforward assessment of the sensitivity as a function of different parameters, such as the fiducial mass. It does not, however, consider background uncertainties, but assumes perfect knowledge of the background rates.  

\subsection{Frequentist profile-likelihood analysis}
\label{sec:Sensitivity_Freq}

To account for and effectively constrain the background uncertainties, we apply a profile-likelihood analysis based on the background model discussed in Sect.~\ref{sec:analysis_BackgroundResults_Model}. The inserted signal is a Gaussian peak with $Q_{\beta\beta}$ and $\sigma_E(Q_{\beta\beta})$ according to Eq.~\eqref{eq:E_resolution}, which is scaled by the $^{136}$Xe atoms in the target volume, an activity corresponding to $T_{1/2}^{0\nu}$ and the detection efficiency, as shown in Fig.~\ref{fig:background_spectrum_material}. 

Background uncertainties from the model are treated as nuisance parameters  with Gaussian constraining terms in the likelihood. For external background contributions, their variances are obtained either by the model fit on the spectrum corresponding to $\SI{5}{t}$ FV ($^{208}$Tl peak intensity and slope parameter) or extrapolation of the model parameters from larger fiducial volumes into the low fiducial mass range  ($^{214}$Bi / $^{208}$Tl peak ratio, $^{208}$Tl continuum / $^{208}$Tl peak intensity). The uncertainty on the subdominant contribution from $^{44}$Sc has been neglected. For the intrinsic contributions, the variances correspond to the square of the errors listed in Table~\ref{tab:backgroundsources}. The corresponding slope uncertainties are negligible.

We obtain a $T_{1/2}^{0\nu}$ sensitivity limit of \SI{2.4e27}{yr} for a 10 year exposure with \SI{5}{t} fiducial mass. The corresponding $3\,\sigma$ discovery potential after 10 years exposure is \SI{1.1e27}{yr}.

\section{Discussion}
\label{sec:summary}
The DARWIN observatory will reach a sensitivity to the neutrinoless double beta decay of $^{136}$Xe of \SI{2.4e27}{} years $T_{1/2}$ exclusion limit (90\%~C.L.) and a discovery sensitivity ($3\,\sigma$) of $T_{1/2} = \SI{1.1e27}{}$ years after 10 years of exposure. 

In the baseline scenario discussed above, the assumptions on radio-purity and detector performance are considered realistic or even conservative. In an optimistic scenario, the external background could be reduced by a factor of three or more. The required measures include the use of less radioactive PMTs (with reduced mass of ceramic \cite{Aprile:2015lha}) and/or low radioactivity SiPMs, more stringent material selection to reach lower levels of radio-activity for PTFE~\cite{Auger:2012gs}, copper~\cite{Algrall:2016tmd} and titanium, as well as  more radio-pure electronics.

Intrinsic backgrounds, dominated by the muon-induced activation of $^{136}$Xe, are difficult to mitigate assuming the muon flux at \SI{3500}{} meter water equivalent (mwe) depth of LNGS. A time- and spatial- muon veto might allow for suppression by up to a factor of two at an acceptable exposure loss. The $^{137}$Xe contribution would, however, become subdominant in a sufficiently deep laboratory. A total intrinsic background suppression by a factor of five or even eight could then be reached assuming a reduced BiPo tagging inefficiency of $0.1\%$  and $0.01\%$, respectively. Assuming a factor five reduction in external sources the latter scenario leads to a  solar $^8$B neutrino dominated background.

The sensitivity could be increased by further exploitation of the SS/MS discrimination, discussed in Sect.~\ref{sec:signal}. 
Despite increased signal rejection, the gain in background reduction dominates for spatial separation thresholds down to $\epsilon = \SI{3}{mm}$.
The cluster separation in the $x$-$y$-plane would benefit from a higher granularity photosensor top array, featuring e.g. SiPMs. The $z$-position reconstruction is already more accurate and a combined three dimensional charge signal analysis will optimize the separation.

\begin{figure}[htb]
\begin{center}
\includegraphics[width=\linewidth]{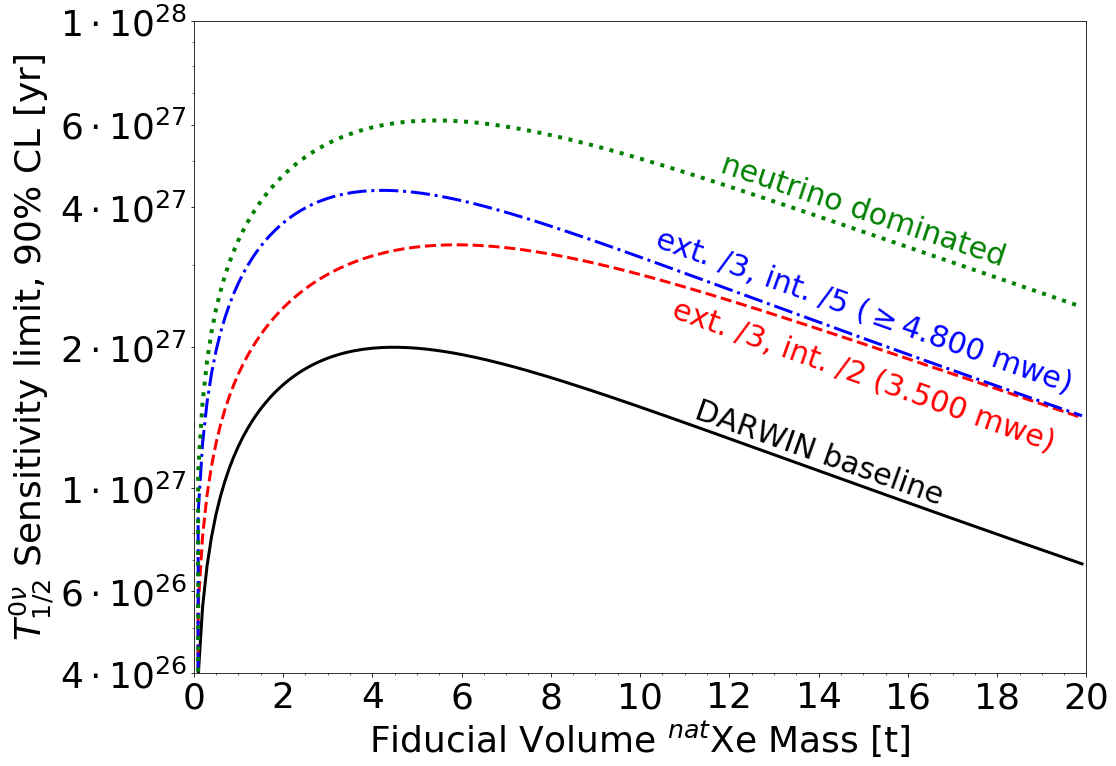}
\includegraphics[width=\linewidth]{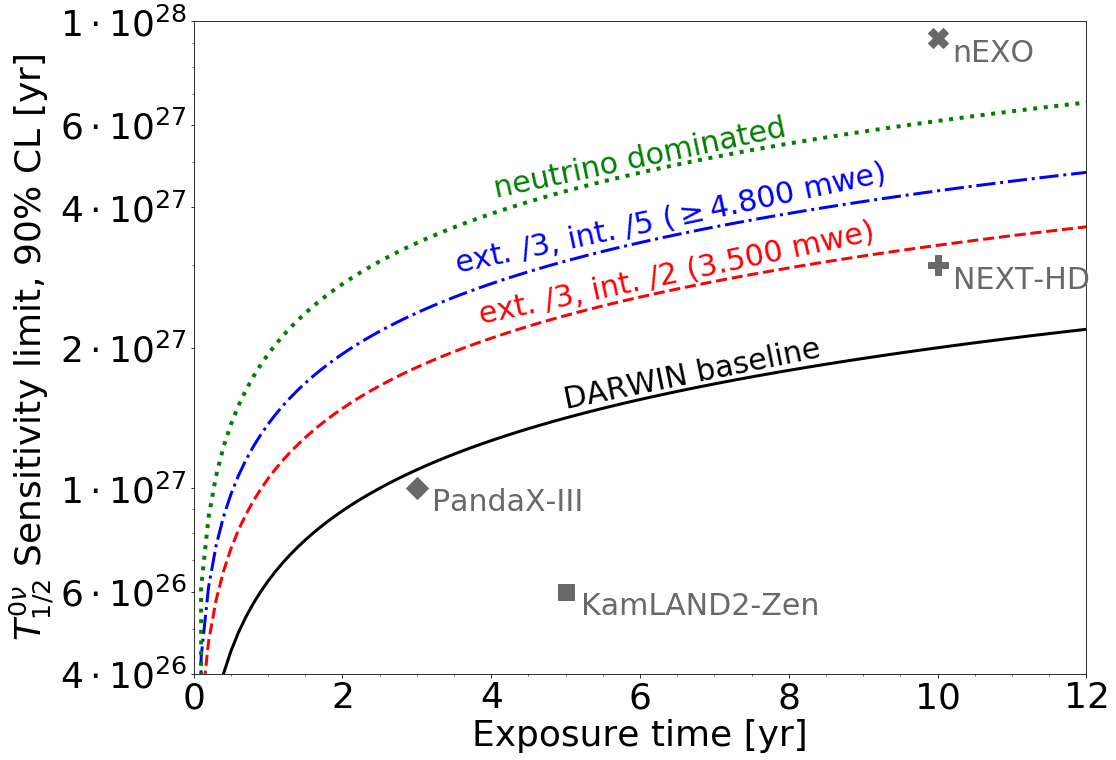}
\caption{DARWIN median $T_{1/2}^{0\nu}$ sensitivity at 90\%~C.L. as a function of {\color{black}fiducial volume mass for 10 years of exposure (top) as well as of} the exposure time {\color{black}for the optimized fiducial volume (bottom)}. The baseline design is compared with different optimistic scenarios. The latter assume a reduction of the external (ext.) and the intrinsic (int.) backgrounds and improved spatial separation threshold of $\SI{10}{mm}$ (red, blue) or $\SI{5}{mm}$ (green). Sensitivity projections for future $^{136}$Xe $0\nu\beta\beta$ experiments are shown for comparison~\cite{Gomez_NEXT:2019,2017SCPMA..60f1011C,Albert:2017hjq,Barabash:2019zas}.}
\label{fig:Sensitivity_Exposure}
\end{center}
\end{figure}

\begin{figure}[ht]
\begin{center}
\includegraphics[width=0.95\linewidth]{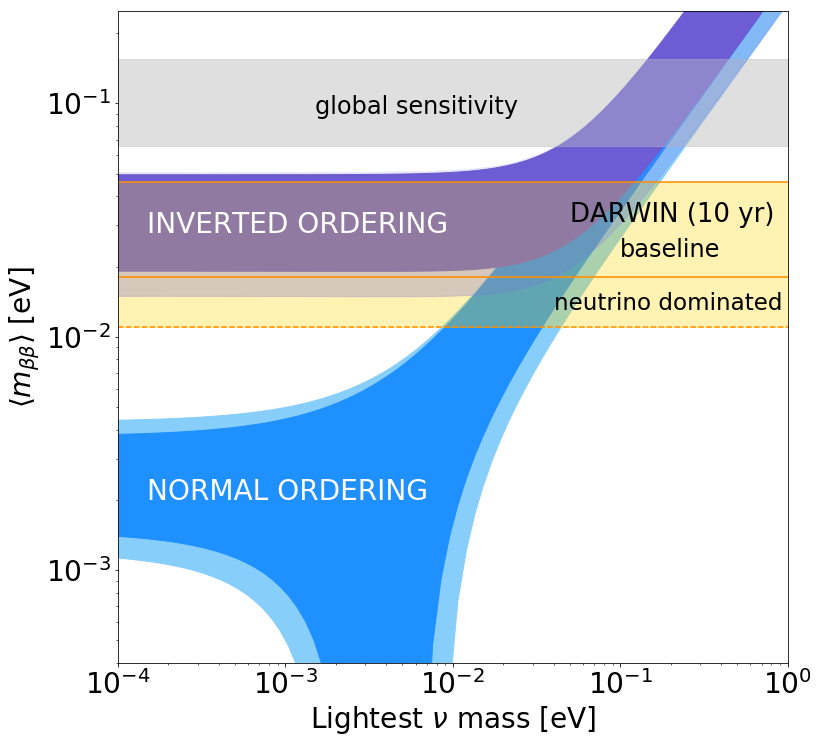}
\caption{Effective Majorana neutrino mass vs. lightest neutrino mass. The sensitivity reach after 50\,t$\times$yr of  exposure is shown for the baseline and the optimistic neutrino dominated scenario. The horizontal bands stem from the range of nuclear matrix elements~\cite{Engel:2016xgb}. Global sensitivity according to~\cite{Agostini:2019gre}, oscillation parameters from~\cite{Esteban:2018azc,NuFit:2019}.}
\label{fig:MajoranaMass}
\end{center}
\end{figure}

The largest sensitivity increase can be achieved with a combination of the above mentioned measures. Fig.~\ref{fig:Sensitivity_Exposure} shows the {\color{black}fiducial volume mass dependency (top) and} time evolution {\color{black}(bottom)} of the DARWIN half-life limit sensitivity (90\%~C.L.) calculated with the figure-of-merit estimator (see Sect.~\ref{sec:Sensitivity_F-o-M}) for the baseline and different optimistic scenarios. {\color{black}The latter assume} reduced spatial separation threshold $\epsilon$, intrinsic and external background rates. Fig.~\ref{fig:MajoranaMass} translates the half-life limit sensitivity to the effective Majorana neutrino mass $m_{\beta\beta}$ using Eq.~(\ref{eq:Majorana_mass}), where the $m_{\beta\beta}$ range corresponds to the range of published nuclear matrix elements~\cite{Engel:2016xgb}. Under the conservative baseline assumptions, DARWIN reaches a $m_{\beta\beta}$ limit of [18-\SI{46}{meV}].
The neutrino dominated scenario yields a limit in the [11-\SI{28}{meV}] range. 
{\color{black}Future dedicated neutrinoless double beta decay experiments using either $^{136}$Xe or other isotopes are aiming for a similar science reach as DARWIN, as shown for comparison in Table~\ref{tab:Sensitivities} and in Fig.~\ref{fig:Sensitivity_Exposure} (bottom). }

\begin{table*}[!t]
\begin{center}
\begin{tabular}{l c c c c c}
\hline
Experiment              & Isotope   & \multicolumn{2}{c}{ Sensitivity limit (90\% C.L.)} & Exposure time & Reference \\
                        &           & $T_{1/2}^{0\nu}$  [yr]          & $m_{\beta\beta}$ [meV] & [yr] & \\
\hline
DARWIN (baseline)       & $^{136}$Xe    & \SI{2.4e27}{}     & 18-46 & 10 & this work \\
DARWIN ($\nu$ dominated)  & $^{136}$Xe    & \SI{6.2e27}{}   & 11-28 & 10 & this work  \\
\hline
KamLAND2-Zen            & $^{136}$Xe    & \SI{6e26}{}       & 37-91 & 5  & ~\cite{Barabash:2019zas} \\
PandaX-III              & $^{136}$Xe    & \SI{1e27}{}       & 28-71 & 3 & ~\cite{2017SCPMA..60f1011C} \\
NEXT-HD                 & $^{136}$Xe    & \SI{3e27}{}       & 16-41 & 10 & ~\cite{Gomez_NEXT:2019} \\
nEXO                    & $^{136}$Xe    & \SI{9.2e27}{}     & 9-23 & 10 & ~\cite{Albert:2017hjq} \\
\hline
SNO+-II                 & $^{130}$Te    &  \SI{7e26}{}      & 20-70 & 5 & ~\cite{Barabash:2019zas} \\
AMoRE-II                & $^{100}$Mo    &  \SI{5e26}{}      & 15-30 & 5 & ~\cite{Barabash:2019zas} \\
CUPID                   & $^{130}$Te~/~$^{100}$Mo & (2-5)$\times 10^{27}$ & 6-17 & 10 & ~\cite{Barabash:2019zas}\\
LEGEND-1000             & $^{76}$Ge     & \SI{1e28}{} & 11-\SI{28}{} & 10 & ~\cite{Barabash:2019zas} \\
\hline
\end{tabular}
\captionsetup{width=0.85\textwidth}
\caption{{\color{black} Comparison of $T_{1/2}^{0\nu}$ and $m_{\beta\beta}$ sensitivity limits (90\% C.L.) between DARWIN and future $0\nu\beta\beta$ experiments. For experiments using $^{136}$Xe the $m_{\beta\beta}$ ranges are calculated with the nuclear matrix element ranges from~\cite{Engel:2016xgb}, those using other isotopes are taken from~\cite{Barabash:2019zas}.}}
\label{tab:Sensitivities}
\end{center} 
\end{table*}

The objective of detecting particle dark matter with a sensitivity down to the neutrino floor requires the DARWIN observatory to be an ultra-low background experiment. It additionally features a high $^{136}$Xe target mass, excellent energy resolution and single site discrimination capability. In the presented baseline scenario DARWIN will reach {\color{black}a sensitivity that approaches that of the tonne-scale proposed $0\nu\beta\beta$ experiments.} Under more optimistic assumptions, requiring adaptations to the baseline design, DARWIN will explore the full inverted hierarchy and will compete with the most ambitious proposed $0\nu\beta\beta$ projects. 

\section*{Acknowledgements}
This work was supported 
by the Swiss National Science Foundation under grants No~200020-162501 and No~200020-175863, 
by the European Union's Horizon 2020 research and innovation programme under the Marie Sklodowska-Curie grant agreements No~674896, No~690575 and No~691164, 
by the European Research Council (ERC) grant agreements No~742789 ({\sc{Xenoscope}}) and No~724320 ({\sc{ULTIMATE}}), 
by the Max-Planck-Gesellschaft, 
by the Deutsche Forschungsgemeinschaft (DFG) under GRK-2149, 
by the US National Science Foundation (NSF) grants No~1719271 and No~1940209, 
by the Portuguese FCT, 
by the Netherlands Organisation for Scientific Research (NWO),
by the Ministry of Education, Science and Technological Development of the Republic of Serbia 
and by grant ST/N000838/1 from Science and Technology Facilities Council~(UK).

\bibliographystyle{JHEP}
\bibliography{DarwinDoubleBeta}

\end{document}